\documentclass[conference]{IEEEtran}
\IEEEoverridecommandlockouts
\usepackage{cite}
\usepackage{amsfonts}
\usepackage{textcomp}
\usepackage{xcolor}
\def\BibTeX{{\rm B\kern-.05em{\sc i\kern-.025em b}\kern-.08em
    T\kern-.1667em\lower.7ex\hbox{E}\kern-.125emX}}
    
\usepackage[final]{graphicx}
\graphicspath{ {./} }
\usepackage[reqno]{amsmath}
\usepackage{amssymb}
\usepackage{amsthm}
\usepackage{epstopdf}
\usepackage{algorithm}
\usepackage{algorithmicx}
\usepackage{algpseudocode}
\usepackage{setspace}
\usepackage[T1]{fontenc}
\usepackage{color,soul}
\usepackage{multirow}
\usepackage{array}
\usepackage{cite}
\usepackage{verbatim}
\usepackage{enumitem}
\usepackage{url}
\usepackage{csvsimple}
\usepackage{float}

\usepackage{float}
\usepackage{graphicx}
\usepackage[capitalize]{cleveref}
\usepackage{environ}

\ifCLASSOPTIONcompsoc
    \usepackage[caption=false, font=normalsize, labelfont=sf, textfont=sf]{subfig}
\else
\usepackage[caption=false, font=footnotesize]{subfig}

\usepackage{array}
\newcommand{\PreserveBackslash}[1]{\let\temp=\\#1\let\\=\temp}
\newcolumntype{C}[1]{>{\PreserveBackslash\centering}p{#1}}
\newcolumntype{R}[1]{>{\PreserveBackslash\raggedleft}p{#1}}
\newcolumntype{L}[1]{>{\PreserveBackslash\raggedright}p{#1}}

\begin{document}

\title{Developing a Transferable Federated Network Intrusion Detection System}

\author{\IEEEauthorblockN{Abu Shafin Mohammad Mahdee Jameel\IEEEauthorrefmark{1}, Shreya Ghosh\IEEEauthorrefmark{1}, Aly El Gamal\IEEEauthorrefmark{1} }
\IEEEauthorblockA{\IEEEauthorrefmark{1} School of Electrical and Computer Engineering, Purdue University, USA}
\IEEEauthorblockA{Email: {\{amahdeej, ghosh64, elgamala\}@purdue.edu}} \vspace{-20pt}
\thanks{\vspace{-20pt} The first two authors contributed equally to this paper.}
}

\maketitle

\begin{abstract}
Intrusion Detection Systems (IDS) are a vital part of a network-connected device. In this paper, we develop a deep learning based intrusion detection system that is deployed in a distributed setup across devices connected to a network. Our aim is to better equip deep learning models against unknown attacks using knowledge from known attacks. To this end, we develop algorithms to maximize the number of transferability relationships. We propose a Convolutional Neural Network (CNN) model, along with two algorithms that maximize the number of relationships observed. One is a two step data pre-processing stage, and the other is a Block-Based Smart Aggregation (BBSA) algorithm. The proposed system succeeds in achieving superior transferability performance while maintaining impressive local detection rates. We also show that our method is generalizable, exhibiting transferability potential across datasets and even with different backbones. The code for this work can be found at https://github.com/ghosh64/tabfidsv2.


\end{abstract}
\IEEEpeerreviewmaketitle
\vspace{-10pt}
\section{Introduction}


Intrusion Detection Systems (IDS) are essential for protecting network infrastructures from unauthorized access and potential threats by identifying and mitigating the efforts of malicious actors. 
Historically, network intrusion detection systems have relied on traditional algorithms like Naive-Bayes classifiers, Random Forest classifiers, and Support Vector Machines (SVM) for threat detection \cite{lansky2021deep}. Methods like stochastic gradient descent based models, hypergraph based machine learning systems, random forest models with class probability distributions have been shown to produce good intrusion detection performances \cite{sgd_nb, hypergraph}. Statistical models such as Logistic Regression, SVMs, Decision Trees, Random Forests, Multi layer Perceptrons have also been shown to perform well with Feature selection methods \cite{ HFRF}. 

While these methods have been foundational in effective IDS development, recent research highlights the enhanced performance of deep learning approaches and show that they surpass traditional algorithms in detecting network intrusions \cite{barnard2022robust}. The popularity of these deep learning methods stems from their exceptional  performance, and their ability to effectively manage imbalanced datasets, a prevalent issue in network security \cite{fu2022deep}. As a result, deep learning techniques are now considered benchmark standards in the development and evaluation of IDS datasets, playing a pivotal role in driving the evolution of future network security measures \cite{koroniotis2019towards}. 

\begin{figure*}[t]
    \centering
    \includegraphics[width=1.0\textwidth]{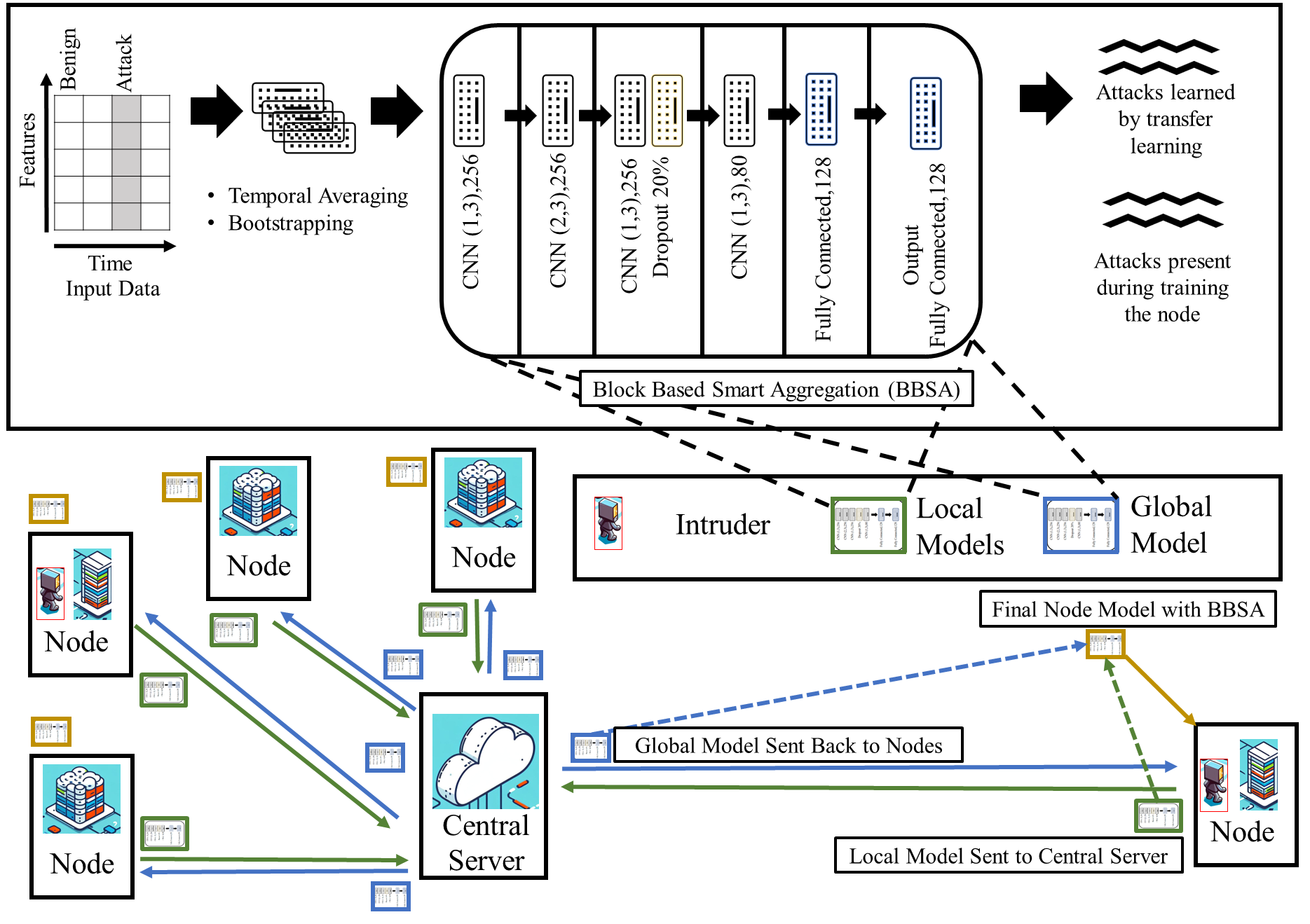}
    \caption{Architecture of proposed intrusion detection model. \vspace{-15pt}}
    \label{fig:sysarch}
    \vspace{-15pt}
\end{figure*}

The surge in research within this area has resulted in the creation of numerous network traffic datasets, such as KDD 99\cite{ghorbani}, NSLKDD\cite{ghorbani}, CIC-IDS 2017\cite{sharafaldin2018toward}, and ToN-IoT\cite{toniot}. These datasets, often compiled using traffic monitoring tools like Wireshark, provide an extensive overview of network activities. However, their complex, high-dimensional nature presents a significant challenge for machine learning models in extracting meaningful information.

Contemporary deep learning models in intrusion detection are often trained to identify specific, previously known attacks or to learn from historical attack patterns. These models excel in detecting familiar threats with high accuracy. However, their capability significantly decreases when confronting novel or zero-day attacks\textemdash unseen and undocumented threats appearing in real-time. The unpredictable nature of these attacks represents a substantial challenge in network security, given the difficulty in predicting the spectrum of possible intrusions. Moreover, anomaly detection approaches tend to categorize all unknown network intrusions under a single class,  thereby failing to offer a comprehensive analysis of model transferability.

The concept of model transferability has gained traction, positing that models could recognize a broader array of attacks beyond their initial training scope. Research into transferability, such as highlighted in \cite{verkerken2022towards} and \cite{catillo2022transferability}, explores this potential. In \cite{verkerken2022towards}, models are trained on a specific dataset and evaluated on another, focusing on a single type of attack. The work presented in \cite{catillo2022transferability} extends this approach by investigating the model's ability to generalize across different attack types, training on one and testing on another. However, it is worth noting that both studies are performed on pre-selected training and testing attack classes.

Recent advances have highlighted a critical challenge traditionally linked to deep learning: the substantial computational power and memory it requires \cite{rizvi2023deep}. Deep learning models are now embedded directly into hardware. The concept of model transferability plays a pivotal role here, allowing for the use of smaller datasets while maintaining the model's ability to generalize across various attack vectors. This strategy significantly cuts down on training times and reduces the computational load and memory requirements, which is especially beneficial for deployment on resource-constrained edge devices.

This integration of edge based federated learning into the realm of Intrusion Detection Systems (IDS) signifies a pivotal shift towards distributed, privacy-preserving, and computationally efficient cybersecurity measures \cite{rahman2020internet}. Research, including findings from \cite{rahman2020internet} and \cite{ruzafa2021intrusion, sdnvanet} highlight federated learning's capability to surpass self-learning models and achieve comparable accuracy to centralized systems in IDS applications.  The exploration of model transferability within federated learning is emerging as a key area of interest \cite{zhang2022federated}.

In this paper, we propose a federated deep learning setup aimed at achieving maximum transferability, while preserving in-class IDS performance. The contributions of this work can be outlined as follows:
\begin{enumerate}
    \item We develop a deep learning model that performs well for both traditional and transferable IDS applications. 
    
    \item We introduce two data preprocessing steps before model training-temporal averaging and bootstrapping that boost the transferability performance, and name the resulting algorithm \textbf{T}emporally \textbf{A}veraged \textbf{B}ootstrapped \textbf{F}ederated \textbf{I}ntrusion \textbf{D}etection \textbf{S}ystem (TabFIDS) version 1.
    
    \item We further propose a novel Block-Based Smart Aggregation(BBSA) algorithm that works with FedAvg\cite{fedavg} aggregation to make performance driven decisions about layer-wise weight aggregation in every round of training. This setup is called TabFIDSv2. In our experiments, we find that TabFIDSv2 outperforms other methods in both localized detection and transferability, while being generalizable across different neural network backbones, and datasets.

    \item We develop a Data-Driven Feature Elimination(DDFE) algorithm that helps us determine which features are most important for model learning.

    
\end{enumerate}
We note that part of the work included in this paper has been presented in \cite{ghosh2024transfer}. This paper is organized as follows: In Section \ref{systemarch}, we explain our system architecture, and the federated learning setup. Next, we present the deep learning model for intrusion detection. We then explain bootstrapping and temporal averaging, the two pre-processing steps to improve the transferability along with our BBSA algorithm. Finally, in Section \ref{results}, we explain our experimental setup and show that our proposed algorithm significantly improves the transferability in a federated setup, performs well even with different backbones, and shows inter dataset transferability. 

\vspace{-5pt}
\section{System Architecture}\label{systemarch}
In this section, we discuss the overall architecture of our federated system environment, as shown in Fig. \ref{fig:sysarch}. 

\vspace{-5pt}
\subsection{Nodes}
Consider a scenario where several devices with networking capabilities are connected on a network. Each of these devices are referred to as a node. There are attackers within the network injecting harmful data packets with the objective of compromising the security of the nodes. Each node however, is equipped with a deep learning based intrusion detection system that should be able to identify the malicious data packets. Our objective is to create a framework in which the intrusion detection softwares leverage traffic data at each node to detect existing as well as rare attacks.
\vspace{-5pt}

\subsection{Intrusion Detection System}

The IDS on each node is trained before deployment on labeled benign and attack data. There is a central server that participates in the training of the IDS, and it may have one of two different roles:

\subsubsection{Centralized Learning}

In centralized learning configuration, the data on each node is sent to a central server where a central model is trained using all the aggregated data. The trained model is sent back to each of the local nodes for deployment and detection. Nodes do not perform any kind of localized training based on the data samples of that node.

\subsubsection{Federated Learning}

In a federated learning configuration, each node trains their local models on local data. Each node then sends their locally trained model to the central server for aggregation. The central server uses an aggregation algorithm to create a single model from the localised models called global model. The global model is then sent back to each of the nodes for another round of continued training. In this regime, there is no data being exchanged over the network, rather only the model weights are sent. 

In our setup, the central server uses a FedAvg \cite{fedavg} aggregation algorithm to aggregate the model parameters sent by the nodes. The global model is sent back to the nodes for further rounds of localised training and this continues over multiple communication rounds. The local and global models have the same architecture as shown in \ref{fig:sysarch} but differ in their parameters. The global model is aggregated from the local models using the following equation:

\begin{equation}
    \forall k, w_{t+1}^k \leftarrow w_t - \eta g_k; w_{t+1} \leftarrow \sum_{k=1}^{K} \frac{n_k}{n} w_{t+1}^k 
\end{equation}
where $w_t$ are the model weights after communication round $t$, $n_k$ is the number of local samples in the training data, $n$ is the total number of data samples across all nodes, assuming we have $K$ nodes in total. 

\subsubsection{Block-Based Smart Aggregation}

\begin{algorithm}[t]
\caption{Block-Based Smart Aggregation (BBSA) Technique}
\label{algorithm:bbsa}
\begin{algorithmic}[1]
\Require $N$ nodes each with local dataset $D_i$, $i \in \{1, 2, \dots, N\}$
\Ensure Optimized global model $G$

\State Initialize global model $G$ with random weights
\For{each communication round $r$}
    \For{each node $i$ in parallel}
        \State Train local model $L_i^r$ on $D_i$
        \State Save local model weights $W_{L_i}^r$ before aggregation
        \State Send local model weights $W_{L_i}^r$ to server
    \EndFor
    \State Aggregate weights on server to update global model $G^r$
    \State Distribute updated global model $G^r$ to all nodes
    \For{each node $i$ in parallel}
        \State Re-train local model $L_i^{r+1}$ on $D_i$ using $G^r$
        \State Save re-trained model weights $W_{L_i}^{r+1}$
        \State \textbf{BBSA Step:} Compare $W_{L_i}^r$ and $W_{L_i}^{r+1}$
        \For{each block $b$ in model}
            \State Select weights for block $b$ from either $W_{L_i}^r$ or $W_{L_i}^{r+1}$ based on performance
        \EndFor
        \State Update $L_i^{r+1}$ with selected block weights
        \State Send updated $L_i^{r+1}$ weights to server for next round
    \EndFor
    \State Aggregate updated weights on server to update $G^{r+1}$
\EndFor
\end{algorithmic}
\end{algorithm}

A notable limitation within the federated averaging methodology is the potential reduction in detection accuracy at the level of individual nodes over successive communication rounds. This challenge arises from the practice of averaging the weights across nodes for updating a model, where a locally optimized set of weights might be more effective for a particular node than the globally aggregated set. 

To mitigate this, a strategy is employed wherein nodes, upon receipt of the global model, undertake a phase of re-training using their local data prior to the deployment of the model, incorporating local specificity with broader, aggregated insights. However, some nodes might still encounter a drop in intrusion detection performance. The diversity in data distribution and attack patterns across nodes can result in varying degrees of effectiveness for this approach, with some nodes potentially not achieving the desired level of accuracy improvement in identifying network intrusions. 

The proposed Block-Based Smart Aggregation (BBSA) technique addresses the challenge of maintaining or enhancing accuracy across individual nodes during iterative communication rounds in a federated learning setup. This method introduces a strategic layer of decision-making in the model updating process that is tailored to optimize performance on a per-node basis.

The algorithm for BBSA is presented in Algorithm \ref{algorithm:bbsa}. Here, nodes save their local model weights prior to dispatching them to the central server for aggregation. Upon receiving the aggregated global model, nodes engage in a re-training session utilizing their specific local datasets. 

Unlike the conventional federated learning process, where this updated model would directly advance to the central server for further aggregation, the BBSA methodology introduces a critical comparative evaluation step. In the proposed selective weight integration process, the decision to utilize weights from either the original local model, as it existed before being sent for aggregation, or the model retrained after receiving the global updates, is made. The primary objective is to harness the strengths of both models—leveraging the specific insights captured by the local model before aggregation and the broader, globally informed updates of the re-trained model. 

The methodology for constructing the combined model involves carefully analyzing the deep learning architecture to identify distinct blocks within the model, delineated based on the function of layers that operate collectively within the architecture. The optimization process entails selecting all the weights within a given block to be exclusively sourced either from the original local model prior to aggregation or from the model that has been re-trained post-aggregation. This block-based selection process is critical for several reasons:

\begin{enumerate}
    \item \textit{Preservation of Functional Integrity:} By treating blocks as the units for weight selection, this approach ensures that the functional integrity of closely interacting layers is preserved. This is crucial because layers within a block often share a contextual relationship, designed to capture specific features or patterns from the input data. 
    \item \textit{Computational Efficiency:} Working with blocks significantly reduces the computational overhead \textemdash instead of making decisions for potentially thousands of individual weights or layers, the process is simplified to a choice between block configurations. This streamlined approach facilitates quicker iterations and optimizations,  especially when computational resources and time are at a premium.
    \item \textit{Enhanced Generalizability:}  By evaluating the effectiveness of entire blocks of layers, the block-based strategy can leverage the collective performance enhancement that these layers provide when their weights are aligned with either the local context (pre-aggregation) or the global learning objective (post-aggregation). This approach allows for a strategic enhancement of the model's ability to generalize across diverse datasets and attack scenarios, potentially leading to better detection capabilities.
\end{enumerate}

In summary, by identifying and optimizing blocks of layers as cohesive units, this strategy ensures the computational tractability of the optimization process, and aims to enhance the model's overall performance by maintaining the functional cohesion of layers that are designed to work together.

\vspace{-3pt}
\subsection{Deep Learning Model}

The proposed model is developed empirically with the aim to not only maximize the classification accuracy, but also to uncover transferability\textemdash when the model is trained with one attack and tested on another attack. This model has been tuned to perform deep feature extraction from high dimensional dataset both for well-represented as well as under-represented classes of data. Fig. \ref{fig:sysarch} shows the architecture of this model. 

\vspace{-3pt}
\subsection{Bootstrapping}

A characteristic of network traffic data is that the number of benign data packets encountered significantly surpasses the number of attack data packets, which poses challenges for deep learning models. A popular solution to this problem is bootstrapping, where the minority class is re-sampled and appended to the majority class until the dataset is balanced.

\vspace{-4pt}
\subsection{Temporal Averaging}

Temporal averaging, when employed as a pre-processing step, has been shown to boost the transferability of an IDS system \cite{ghosh2023an}. Mathematically this can be given as:

\begin{equation}
    y_{t}=\frac{\sum_{i=0}^{r-1} x(t-i)}{r}, r=\text{window size} 
\end{equation}
where $x_{t}$ represents the input data sample at time $t$ and $y_{t}$ represents the corresponding temporally averaged data sample. Thus each input sample to the model is the temporal average of $r$ other samples.  Temporal averaging further improves privacy of the data, as the deep learning model is trained with the pre-processed data and it never sees the original data.

\subsection{Data-Driven Feature Elimination (DDFE)}

\begin{algorithm}[t] 
\caption{Data-Driven Feature Elimination (DDFE)}
\label{algorithm:ddfe}
\begin{algorithmic}[1]
    \Require TabFIDS model, Full 1D input feature vector
    \Ensure Reduced feature set model with maintained accuracy
    
    \State \textbf{Initial Training:}
    \State Train the TabFIDS model on the full 1D input feature vector $\{F: lenght(F) = N\}$
    \For{each feature $f_i$ in the input feature vector }
        \State Create a modified version of the feature vector with $f_i$ set to zero
        \State Train/evaluate the model with the modified feature vector
        \State Record the change in classification accuracy
    \EndFor
    \State \textbf{Performance Assessment:}
    \For{each feature $f_i$}
        \If{removal of $f_i$ does not significantly reduce accuracy}
            \State Mark $f_i$ as non-contributory
        \EndIf
    \EndFor
    \State \textbf{Simplification:}
    \State Create a reduced feature vector $\{\hat{F} :$ \textit{nonzero feature length} $= (N-n)\}$ by setting all $n$ non-contributory features to zero
    \State Deploy TabFIDS model with the reduced feature set
    
\end{algorithmic}
\end{algorithm}

To reduce the computational cost of deep learning systems, optimizing the size of the feature vector is crucial. The Data-Driven Feature Elimination (DDFE) approach is designed to iteratively evaluate the impact of each feature on the model's performance and then eliminate those that contribute minimally. The algorithm for DDFE is presented in Algorithm \ref{algorithm:ddfe}. The DDFE process involves the following steps:

\begin{enumerate}
    \item \textit{Initial Training:} Train the TabFIDS model on a specific node using the entire 1D input feature vector to establish baseline performance metrics.
    \item \textit{Feature Elimination Simulation:} 
    \begin{itemize}
        \item For each feature in the input vector, create a modified version of the input where the current feature's value is set to zero, simulating its removal.
        \item Assess the model's performance with the modified input vector to determine the impact of the feature's removal on classification accuracy.
    \end{itemize}
    \item \textit{Performance Assessment:} Evaluate the model's performance with the removal of each individual feature, focusing on classification accuracy. Identify features whose elimination does not impose a large performance penalty.
    \item \textit{Simplification:} 
    \begin{itemize}
        \item Remove non-contributory features by setting their values to zero in the input vector, resulting in a reduced feature set.
        \item Train or fine-tune the model on this optimized feature set to achieve a balance between computational efficiency and detection accuracy.
    \end{itemize}
\end{enumerate}

By employing the DDFE method, TabFIDS can maintain high detection accuracy while operating within computational resource constraints, making it particularly suitable for environments where processing power and memory are limited.

\section{Experimental Results} \label{results}

In this section, we first present details about the dataset, experimental setup and the evaluation metric. Then we provide a step by step analysis of the development of the proposed TabFIDSv2, with each step highlighting the impact of individual components of the algorithm. Then we analyze the generalization performance of TabFIDSv2 across different deep learning backends, and across datasets.

\subsection{Dataset}

\begin{figure*}[t]
    \centering
    \subfloat[\label{fig:cen_trans}]{%
       \includegraphics[width=0.28\linewidth]{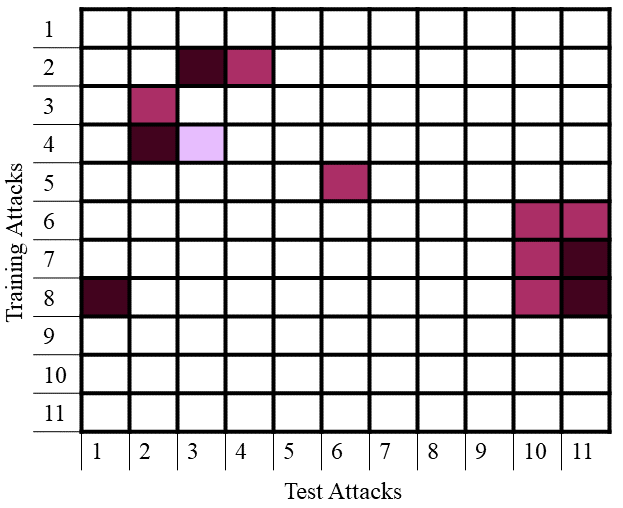}}
    \hfill
    \subfloat[\label{fig:fed_trans_normal}]{%
        \includegraphics[width=0.28\linewidth]{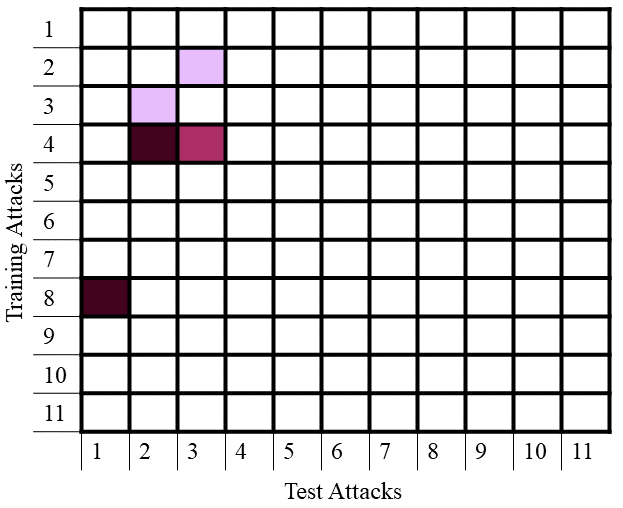}}
    \hfill
    \subfloat[\hspace{68pt}\label{fig:federated_bootstrapping_trans}]{%
        \includegraphics[width=0.28\linewidth]{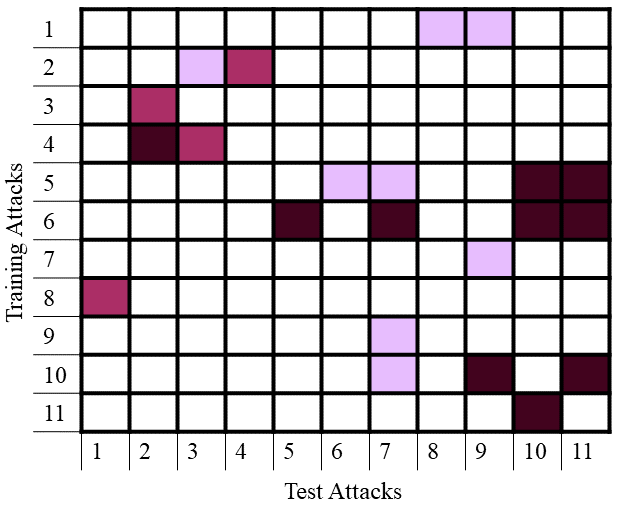}
        \includegraphics[width=0.13\linewidth]{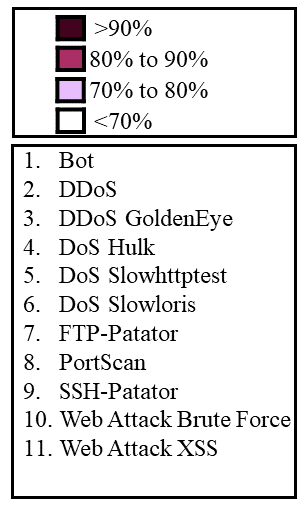}}
    \\
    \vspace{-5pt}
    \subfloat[\label{fig:fed_temporal_trans_3}]{%
       \includegraphics[width=0.28\linewidth]{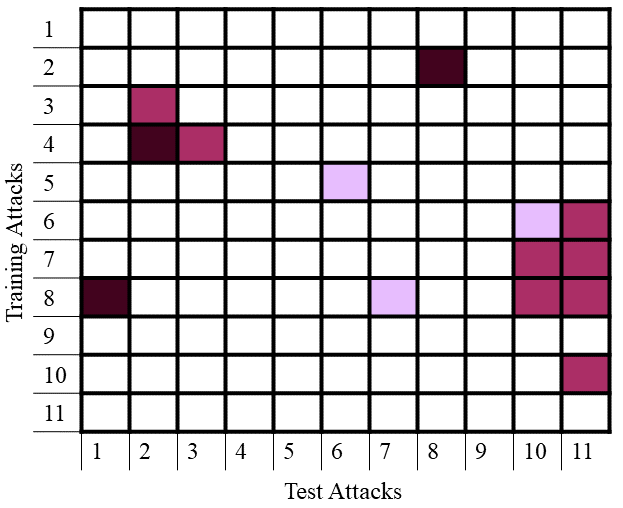}}
    \hfill
    \subfloat[\label{fig:fed_temporal_trans_5}]{%
        \includegraphics[width=0.28\linewidth]{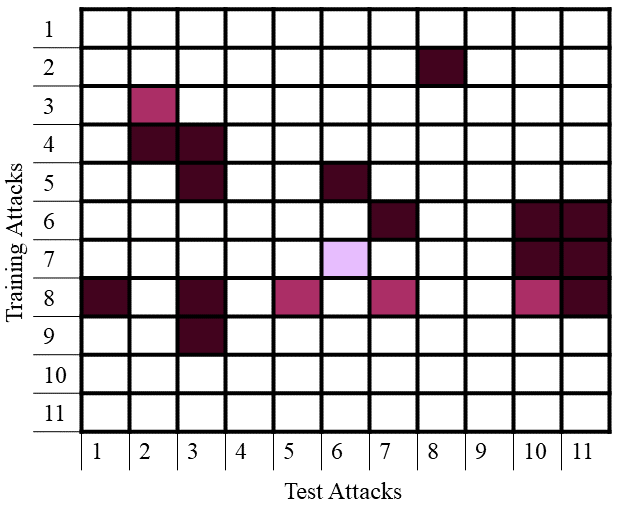}}
    \hfill
    \subfloat[\hspace{68pt}\label{fig:fed_temporal_trans_7}]{%
        \includegraphics[width=0.28\linewidth]{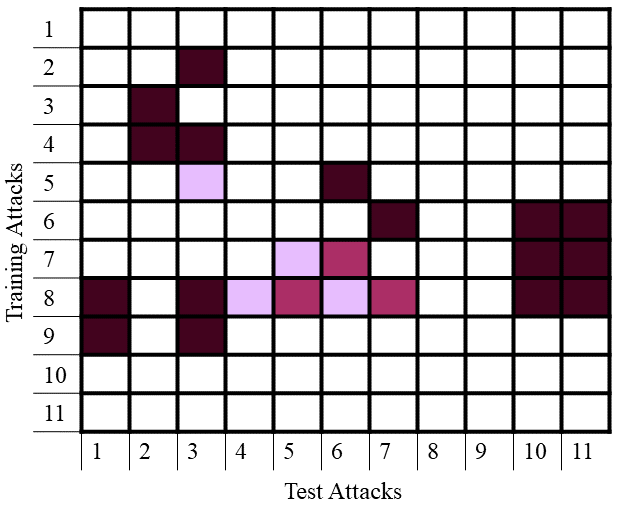}
        \includegraphics[width=0.13\linewidth]{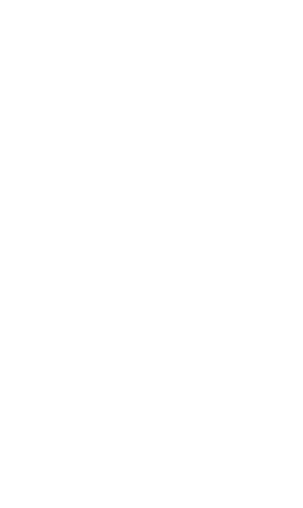}}
    \hfill
    \caption{Transferability Relationships of (Train,Test) Attack Pairs for (a) A Centralized Setup, (b) A Federated Setup, (c) Federated Learning with Bootstrapping, and (d,e,f) Federated Learning with Temporal Averaging (d) Window Length of 3, (e) Window Length of  5, (f) Window Length of 7. \vspace{-15pt}}
    \label{fig:cen_fed_boot_temp_trans}
    \vspace{-5pt}
\end{figure*}

We employ the CIC-IDS 2017 dataset from the Canadian Institute of Cybersecurity \cite{sharafaldin2018toward}. This dataset features 78 features and 14 distinct attack categories. However, the data distribution is very unbalanced: 80\% of it is benign, while the remaining 20\% consists of various forms of attack data. We ignore three attack classes that suffer from extremely low data availability, together constituting only 0.00232\% of total data. Even then, the smallest of the remaining attack categories represents just 0.023\% of the total dataset,  making it extremely challenging for a data-driven IDS. We use 80\% of this data for training, 10\% for validation and 10\% for testing.

\subsection{Evaluation Metric} \label{subsec:evaluation}
Intrusion detection systems typically encounter imbalanced datasets, with significantly more benign data packets than attack data packets. This imbalance poses a unique evaluation challenge, as high overall accuracy may mask poor intrusion detection performance. For instance, in a dataset with 90\% benign data, a classifier achieving 90\% accuracy might fail to detect any attack packets.

To address this problem, precision and recall metrics are often reported in addition to accuracy, providing a more nuanced understanding of the system's performance. Let $tp$, $tn$, $fp$ and $fn$ represent accurately detected attack data points, accurately detected benign data points, wrongly detected attack data points, and wrongly detected benign data points, respectively. The overall accuracy of the system would then be calculated as $(tp+tn)/(tp+tn+fp+fn)$. Precision is $(tp)/(tp+fp)$ and recall would be $(tp)/(tp+fn)$. The recall value provides the ratio of attack data samples that are accurately classified. However, it does not contain information about whether the system can accurately classify benign data samples. 

To overcome this issue, we propose to use the attack accuracy, defined as:

\begin{equation}
    Accr=\frac{\frac{tn}{tn+fp}+\frac{tp}{tp+fn}}{2}
\end{equation}
which gives equal weight to correct detection of benign and attack data packets. In this paper, we use the attack accuracy metric to quantify the performance of IDS systems in our severely imbalanced dataset.

\subsection{Transferability in a centralized setup}

In the next few subsections, we will gradually introduce individual components of the proposed TabFIDSv2, with accompanying analysis of the design choices. 

\subsection{Experimental Setup}

In a centralized environment, the  model is trained on all available benign data in the training set and one class of attack data. The transferability is then evaluated by testing the model against test data packets from attack classes it has not seen during the training phase. We perform this experiment for all 11 attack classes.

On the other hand, in a federated setup, we divide the training dataset to 11 separate nodes. During training, we expose each node to a single attack type only. There is no overlap between the benign data present in different nodes during training. Each node only sees a $\frac{1}{11}$ fraction of the benign data and one attack class, effectively making local training data set availability much lower compared to a centralized system. 

When employing bootstrapping, we sample attack data packets from the same class with replacement until the dataset contains 50\% benign data and 50\% attack data. 
We employ the Adam\cite{adam} optimizer with a learning rate of 0.001 and the loss function is CrossEntropy Loss. For the federated model, we run 20 rounds of federated model aggregation.

For the purpose of this study, we define transferability as an IDS achieving an attack accuracy $\geq$70\% on a test attack class after training on a different attack class. Transferability pairs are denoted by (training attack, test attack) and categorized by performance: very high ($\geq$90\%), high (80-90\%), or moderate (70-80\%) transferability.

\subsection{Centralized vs Federated Transferability}

We start our analysis with an investigation of the transferability relationships for the centralized and federated approach. First, in Fig. \ref{fig:cen_fed_boot_temp_trans}(a), we observe that the centralized model performs reasonably well, identifying 13 transferable pairs. In contrast, the federated approaches yield disappointing results, with only 5 transferable pairs detected, as presented in Fig. \ref{fig:cen_fed_boot_temp_trans}(b). This is surprising, given that federated learning facilitates information exchange between nodes with diverse training attacks during aggregation, which should enhance transferability. However, the smaller individual training sets at each node, combined with the imbalanced dataset, may contribute to the significant decline in transferability performance.

We then employ bootstrapping as a pre-processing step to mitigate the impact of the imbalanced dataset in the federated setup. We present the results in Fig. \ref{fig:cen_fed_boot_temp_trans}(c). This augmentation leads to a substantial improvement in performance, with the detection of 22 transferable pairs, a notable increase from the 5 pairs detected without bootstrapping.

\subsection{Temporal Averaging}

\begin{figure*}[t]
    \centering
    \subfloat[\label{fig:tabfidsv1_federated_80}]{%
       \includegraphics[width=0.28\linewidth]{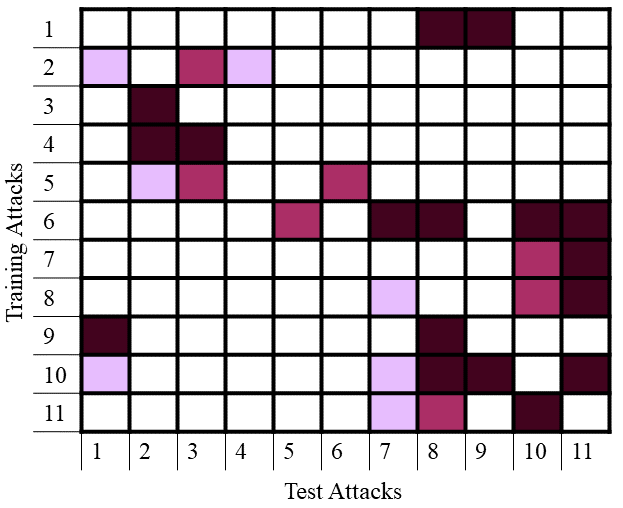}}
    \hfill
    \subfloat[\label{fig:tabfids_central}]{%
        \includegraphics[width=0.28\linewidth]{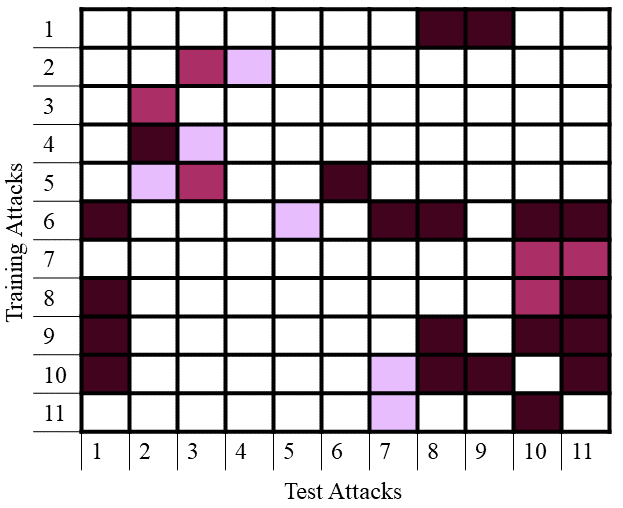}}
    \hfill
    \subfloat[\hspace{68pt}\label{fig:tabfidsv1_80_DDFE_gt}]{%
        \includegraphics[width=0.28\linewidth]{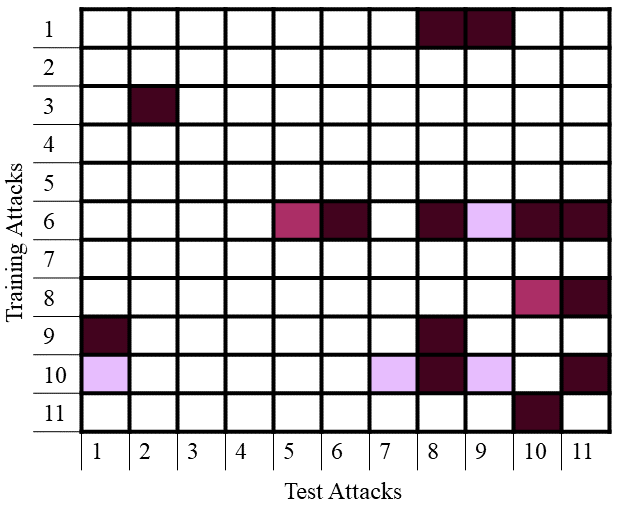}
        \includegraphics[width=0.13\linewidth]{CICIDS_Legend.PNG}}
    \hfill
    \\
    \vspace{-5pt}
    \subfloat[\label{fig:tabfidsv2_80}]{%
       \includegraphics[width=0.28\linewidth]{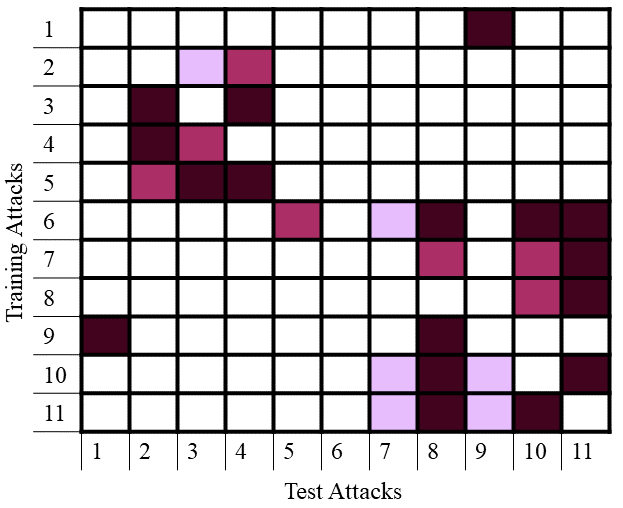}}
    \hfill
    \subfloat[\label{fig:tabfidsv2_50}]{%
        \includegraphics[width=0.28\linewidth]{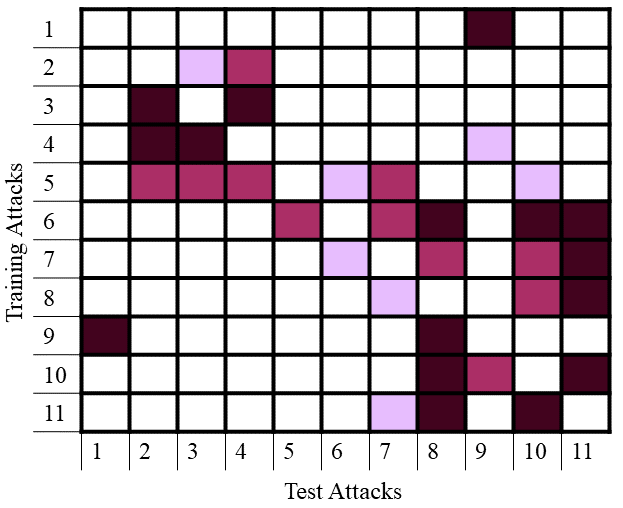}}
    \hfill
    \subfloat[\hspace{68pt}\label{fig:tabfidsv2_25}]{%
        \includegraphics[width=0.28\linewidth]{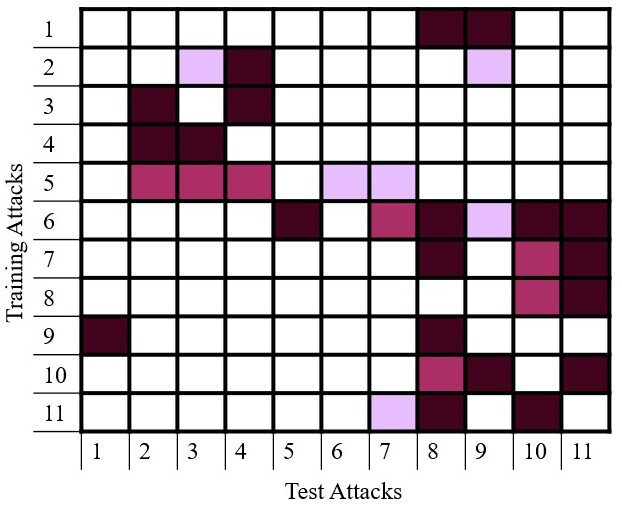}
        \includegraphics[width=0.13\linewidth]{CICIDS_Legend_plcholder.PNG}}
    \hfill
    \caption{Transferability Relationships of (Train,Test) Attack Pairs for (a) TabFIDSv1 - Federated Setup with Bootstrapping and Temporal Averaging, (b) Centralized Setup with Bootstrapping and Temporal Averaging, (c) TabFIDSv1 with Data-Driven Feature Elimination, and (d,e,f) TabFIDSv2 - Block-Based Smart Aggregation combined with TabFIDSv1, with a Train-Validation-Test Split of (d) 80-10-10, (e) 50-10-10, and (f) 25-10-10. }
    \label{fig:Tabfids_v1_v2}
    \vspace{-15pt}
\end{figure*}

We investigate the effect of temporal averaging on transferability, with the results in Fig. \ref{fig:cen_fed_boot_temp_trans}(d) demonstrating a remarkable improvement over the federated setup, with 14 transferable pairs detected compared to 5. Here we average three data packets and feed them to the input layer. This enhancement is noteworthy, as it demonstrates that temporal averaging can boost deep learning model performance even in the presence of severe dataset imbalance, without relying on data augmentation. This can be attributed to the ability of temporal averaging to aggregate multiple input data packets, providing the IDS with a more comprehensive understanding of the network environment. This approach is particularly valuable in resource-constrained training scenarios, where data augmentation can substantially increase training time.

\subsubsection{Window Length for Temporal Averaging}

One question arising from the previous discovery is the impact of the number of samples that are averaged before the input layer. We define this as the window length of temporal averaging. In Fig. \ref{fig:cen_fed_boot_temp_trans}(e) and Fig. \ref{fig:cen_fed_boot_temp_trans}(f), we present the results for window lengths of 5 and 7, respectively. The results illustrate that there is a correlation between the window length and the number of transferable pairs. A window length of five results in an increase to 19 transferable pairs, and a window length of seven gives us 23 transferable pairs, which is higher than the bootstrapping approach. Thus with a simple pre-processing operation we can achieve comparable trasferability to bootstrapping while avoiding the massive increase in training dataset size. The tradeoff here is the introduction of a latency even during deployment in a live system. A bigger window requires more input data packets, resulting in a higher latency penalty.

\subsection{TabFIDS}

Next, we combine bootstrapping and temporal averaging to create the Temporally Averaged Bootstrapped Federated Intrusion Detection System (TabFIDS). This approach yields the most impressive transferability results yet, achieving transferability for 31 pairs, as seen in Fig. \ref{fig:Tabfids_v1_v2}(a). To minimize the latency penalty we use a temporal averaging window size of three. However if better performance is desired and a small increase in processing time is not an issue, we can deploy the same system with a larger temporal averaging window. As shown in Fig. \ref{fig:Tabfids_v1_v2}(a) and later summarized in Table \ref{table:transferable-pairs}, TabFIDS not only excels in transferability but also achieves the highest number of pairs so far with attack accuracy above 90\% (17 pairs), outperforming the next best approach (bootstrapping only) (10 pairs). Furthermore, TabFIDS demonstrates exceptional performance in localized intrusion detection, attaining an average accuracy of 97.18\% across nodes when tested with the same attack class used during training, showcasing its robustness and effectiveness.

Notably, when combining temporal averaging and bootstrapping in TabFIDS, we observe a trade-off, where 5 transferability pairs detectable by standalone bootstrapping and 2 pairs from standalone temporal averaging approach are no longer detectable. This suggests that refining the integration method could further enhance TabFIDS.

Another finding that is worth investigating is the impact of the TabFIDS pipeline on a centralized setup, illustrated in Fig. \ref{fig:Tabfids_v1_v2}(b). The integration of bootstrapping and temporal averaging in a centralized setup also results in improvements in performance, resulting in 32 transferable pairs. This highlights the versatility and broad applicability of the proposed approach.

\subsection*{Impact of DDFE}


Next, we implement the DDFE algorithm mentioned in Algorithm \ref{algorithm:ddfe}, along with TabFIDS. The results are presented in Fig. \ref{fig:Tabfids_v1_v2}(c). DDFE's ability to judiciously reduce the feature set by 39.34\% on average\textemdash ranging from a minimal 8.97\% reduction for specific attack types to as much as 74.36\% for others\textemdash underscores its effectiveness. DDFE retains 19 transferable pairs (38.7\% reduction in transferable pairs), with 13 pairs achieving high transferability rates (>90\%). This demonstrates DDFE's effective feature selection, preserving the model's ability to generalize across unseen attacks. We observe that the introduction of DDFE decreases transferability performance the most for pairs which previously had moderate transferability. DDFE emerges as a valuable tool in resource-constrained setups, where balancing speed and performance is crucial, and a reduction in transferability can be tolerated with a comparable reduction in computational complexity.

\subsection{TabFIDS v2: Incorporating TabFIDS and BBSA}

In Fig. \ref{fig:Tabfids_v1_v2}(d), we present the transferable pairs achieved by TabFIDSv2, an enhanced version of TabFIDS incorporating block-based smart aggregation. Interestingly, the introduction of the BBSA algorithm results in a reduction in the number of transferable pairs, from 31 to 30. However for localized intrusion detection (training and testings sets from the same attack class), performance improves from 97.18\% to 99.64\%.

\subsection{Fine Tuning TabFIDSv2: Size of Training Dataset}

We cannot directly train for transferability, as the unknown attack classes are not available during training. Moreover, the BBSA algorithm focuses on optimizing localized intrusion detection through recursive fine-tuning of neural architecture blocks, and this may compromise transferability. We suspect overfitting to training classes causes this issue. 

To test this, we reduce the size of the available training dataset. Up to the previous section, the train-validation-test  split was 80-10-10. In this experiment, we only use 50\% of the available data for training, instead of 80\%. For the purpose of consistency, the validation and test sets remain unchanged, with 10\% of the data in each. The remaining 30\% of data is not used. The results are presented in Fig. \ref{fig:Tabfids_v1_v2}(e). 

Here, we observe an interesting trend. Reducing the size of the training set improves transferability (we get 34 transferable pairs compared to 30 with a 80-10-10 split) while having a very small decrease in localized intrusion detection performance (from 99.64\% to 99.45\%). This shows that a reduction in the training set has an advantageous impact by making the training more transferable.

Next, we test a 25-10-10 train-validation-test split, with the results presented in Fig. \ref{fig:Tabfids_v1_v2}(f). We get 33 transferable pairs. While the total number of transferable pairs is one less than the 50-10-10 case, on a closer look we notice that here 20 pairs showcase a transferablity performance higher than 90\% (dark purple), compared to 16 pairs in the previous case. In the 25-10-10 case, localized intrusion detection performance decreases slightly, to 98.09\%. In our opinion, This presents a good trade off between strong transferability and good localized performance.

\begin{figure}[t]
    \centering
    \includegraphics[width=0.45\textwidth]{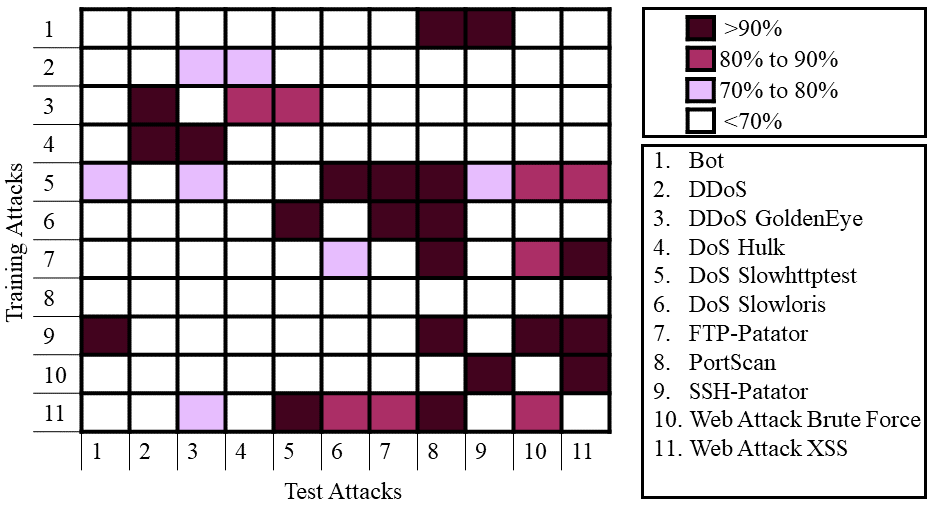}
    \caption{Transferability Relationships for TabFIDSv2 with a Train-Validation-Test split of 05-10-10.}
    \label{fig:tabfidsv2_05}
\end{figure}

Finally, we also perform tests using a 5-10-10 split, where only 5\% of data is used for training, and the results in Fig. \ref{fig:tabfidsv2_05} show 36 transferable pairs, the highest so far. However, due to the significant reduction in training data size, localized intrusion detection performance suffers and falls to 95.90\%.

The notable insight from this analysis is the finding that the block-based aggregation algorithm in TabFIDSv2 achieves better transferability when aided with a reduction in the training set size. We also note that in addition to facilitating robust and resilient intrusion detection, this reduction in the training set size results in significantly faster training times.

\begin{figure}[t]
    \centering
    \includegraphics[width=0.45\textwidth]{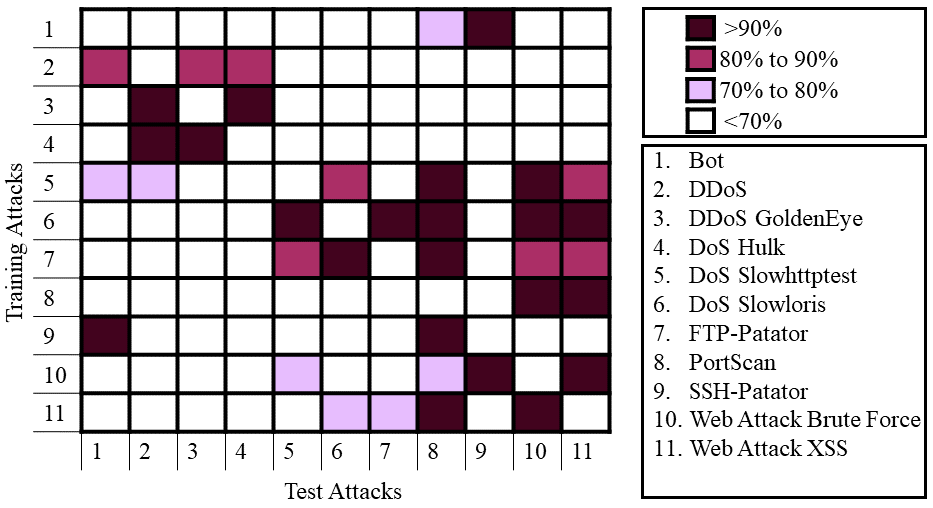}
    \caption{Transferability Relationships for TabFIDSv2 with Temporal Averaging Window of Length 5 (25-10-10 Split).}
    \label{fig:tabfidsv2_temporal_5}
    \vspace{-10pt}
\end{figure}

\begin{figure}[t]
    \centering
    \includegraphics[width=0.45\textwidth]{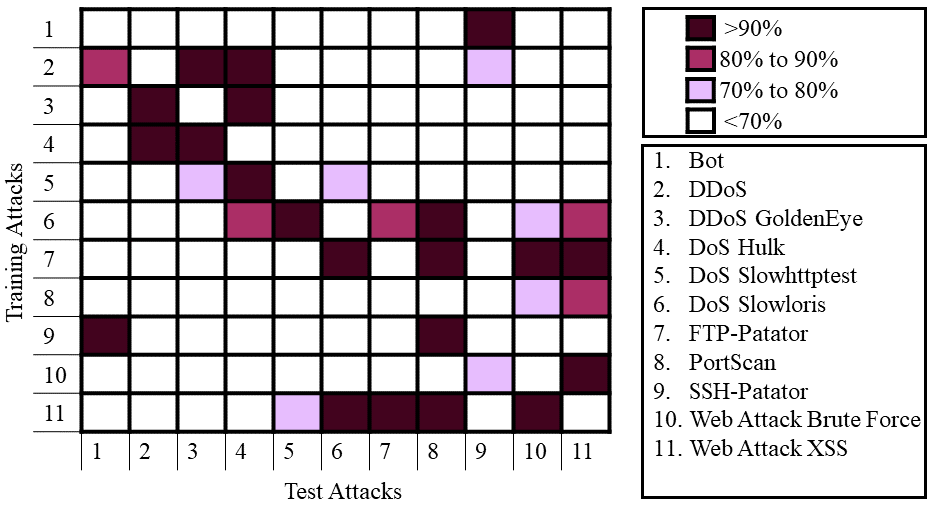}
    \caption{Transferability Relationships for TabFIDSv2 with Temporal Averaging Window of Length 7 (25-10-10 Split).}
    \label{fig:tabfidsv2_temporal_7}
    \vspace{-5pt}
\end{figure}

\begin{table}[tb]
        \centering
        \captionsetup{justification=centering}
	\caption{Number of Occurrences for an Attack Class in Transferable (Train Attack, Test Attack) pairs (TabFIDSv2, 25-10-10 Split, Temporal Averaging Windows of Length 5)}
	\label{table:train-test-stats}
    \tabcolsep=0.12cm
    \begin{tabular}{l|c|c|c|c|c|c|c|c|c|c|c} 
    \hline
     Attack Number& \textbf{1}& \textbf{2}& \textbf{3}&\textbf{4} & \textbf{5}& \textbf{6}& \textbf{7}& \textbf{8}&\textbf{9} & \textbf{10}&\textbf{11}\\
     \hline
     Present as Train Attack& 2& 3& 2& 2& 6& 5& 5& 2& 2& 4& 4\\ 
    \hline
    Present as Test Attack& 3& 3& 2& 2& 3& 3& 2& 7& 2& 5& 5\\
    \hline
    \end{tabular}
    \vspace{-15pt}
\end{table}

\subsubsection{Even Higher Transferability: Window Size}

Next, we focus on extending the analysis on the impact of temporal window size on TabFIDSv2. In Figs. \ref{fig:tabfidsv2_temporal_5} and \ref{fig:tabfidsv2_temporal_7}, we present the impact of temporal windows of sizes 5 and 7 on a 25-10-10 training case. We note that with a temporal window size of 3 in Fig. \ref{fig:Tabfids_v1_v2}(f), we got 33 transferable pairs with a localized intrusion detection of 98.09\%. For a temporal window size of 5, we get 37 transferable pairs (22 with $>90\%$), which is the highest we have achieved. The localized intrusion detection rate also improves to 98.36\%. Interestingly, when we move to a temporal window of 7, the localized intrusion detection rate further improves to 98.82\%, but the number of transferable pairs falls to 33 (20 with $>90\%$). This indicates that there is also an optimal temporal window size for maximum transferablity. 

In Table \ref{table:train-test-stats}, we present an analysis of the (train,test) pairs present in the best performing experiment (TabFIDSv2, 25-10-10 train-validation-test split, temporal windows of size 5). We note that every attack is present in at least two (train,test) pairs, with the DoS Slowhttptest attack showing the highest transferability in detecting other attacks, being present in 6 pairs as train attack. On the other hand, the PortScan attack is the easiest attack to be detected by transferable learning, being present as test attack in 7 pairs.

\subsection{Generalizability of TabFIDSv2: Using a Different Back-end}

TabFIDSv2, as presented above, consists of several components - the federated system architecture, the neural network architecture, temporal averaging, bootstrapping, block-based smart aggregation, and optimal training set selection. So far we have analyzed the individual impacts of all of these components except the neural network architecture. In this section, we analyze the impact of the neural network architecture by replacing the back-end with two different architectures - Resnet and Autoencoder. 

\begin{figure}[t]
    \centering
    \includegraphics[width=0.45\textwidth]{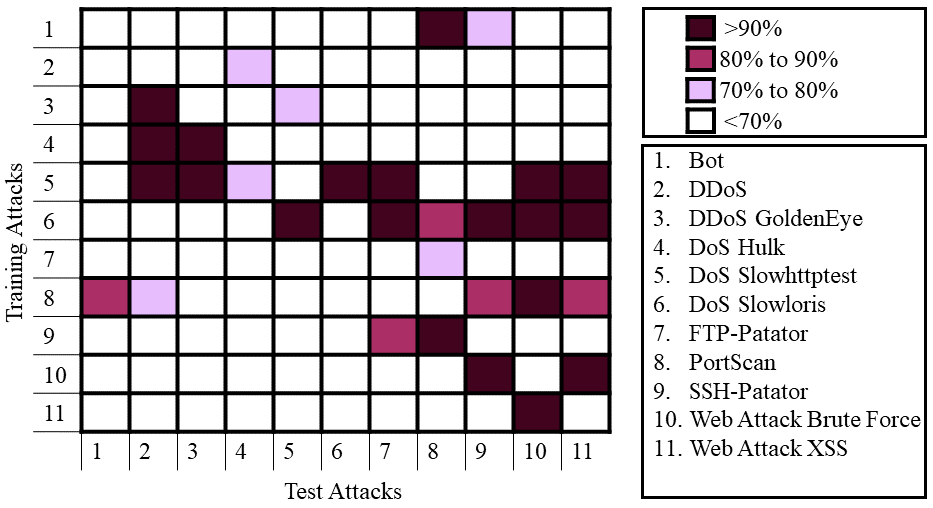}
    \caption{TabFIDSv2 Transferability with a ResNet Backend (25-10-10 Split).}
    \label{fig:resnet_tabfids_v2_25}
    \vspace{-8pt}
\end{figure}

\begin{figure}[t]
    \centering
    \includegraphics[width=0.45\textwidth]{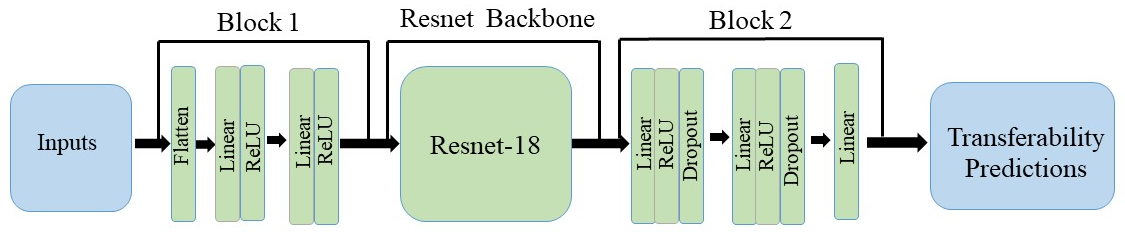}
    \caption{Architecture of the ResNet Backend.}
    \label{fig:resnet architecture}
    \vspace{-15pt}
\end{figure}

\begin{table}[tb]
        \centering
        \captionsetup{justification=centering}
	\caption{Number of Transferable Pairs for Different Approaches and their Attack Accuracy}
	\label{table:transferable-pairs}
    \tabcolsep=0.05cm
    \begin{tabular}{l|c|c|c|c|c} 
    \hline
    & \begin{tabular}[c]{@{}l@{}}Total\end{tabular} & \begin{tabular}[c]{@{}l@{}}>90\%\end{tabular}     & \begin{tabular}[c]{@{}l@{}}80\%\\-90\%\end{tabular}     & \begin{tabular}[c]{@{}l@{}}70\%\\-80\%\end{tabular} & \begin{tabular}[c]{@{}l@{}}Localized\\ Accr \%\end{tabular}          \\ 
    \hline
    Central&       13&   5&      7&      1&    80.09\\ 
    \hline
    Federated&      5&   2&      1&      2&    74.27\\ 
    \hline
    Fed+BStrap&     22&     10&      4&  8&    98.18\\ 
    \hline
    Fed+TempAv(3)&    14&    3&     8& 3&    80.91\\
    \hline
    TabFIDSv1(80/10/10)& 31& 17&    7&    7&    97.18\\
    \hline
    TabFIDSv2(25/10/10)& 33& 20&    7&    6&    98.09\\
    \hline
    TabFIDSv2(25-TempAv 5)& \textbf{37}& 22&    8&    7&    98.36\\
    \hline
    TabFIDSv2(ResNet)& 31& 20&    5&    6&    98.91\\
    \hline
    TabFIDSv2(AutoEnc)& 24& 11&    8&    5&    75.09\\
    \hline
    \end{tabular}
    
\end{table}

\subsubsection{Resnet}

First, we use a resnet architecture presented in Fig. \ref{fig:resnet architecture}, similar to the one in \cite{resnet}. This model consists of 3 blocks. The first block, consisting of linear layers takes the input and resizes it to fit the inputs of the resnet backbone. The resnet backbone is a pretrained 18-layer model, followed by the last block which takes the output from the resnet backbone and classifies it. 

The results from this analysis are presented in Fig. \ref{fig:resnet_tabfids_v2_25}. Here we use a 25-10-10 train-validation-test split. As we can see from the analysis, even with the resnet backend, we find 31 transferable pairs, which shows the generalizability of the proposed approach. This setup achieves a localized intrusion detection rate of 98.91\%.

\subsubsection{Autoencoder} 

\begin{figure}[t]
    \centering
    \includegraphics[width=0.45\textwidth]{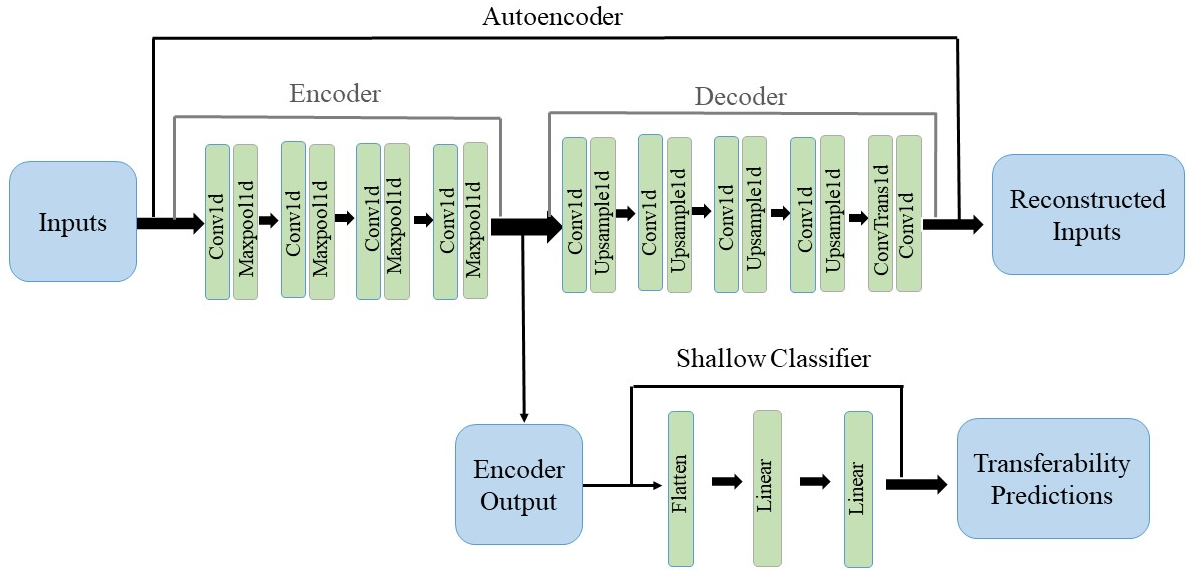}
    \caption{Architecture of the Autoencoder Backend.}
    \label{fig:autoencoder architecture}
    \vspace{-15pt}
\end{figure}

We also utilize an autoencoder based approach, with the architecture\textemdash similar to the approach in \cite{autoenc}\textemdash presented in Fig. \ref{fig:autoencoder architecture}. The autoencoder network consists of an encoder and a decoder. The encoder, using convolutional layers, captures the important features of the network and learns the lower dimensional representation of the input. The decoder consists of convolutional, upsample and transpose convolution layers to reconstruct the original input from the encoded lower dimensional representation. 

\begin{figure}[t]
    \centering
    \includegraphics[width=0.45\textwidth]{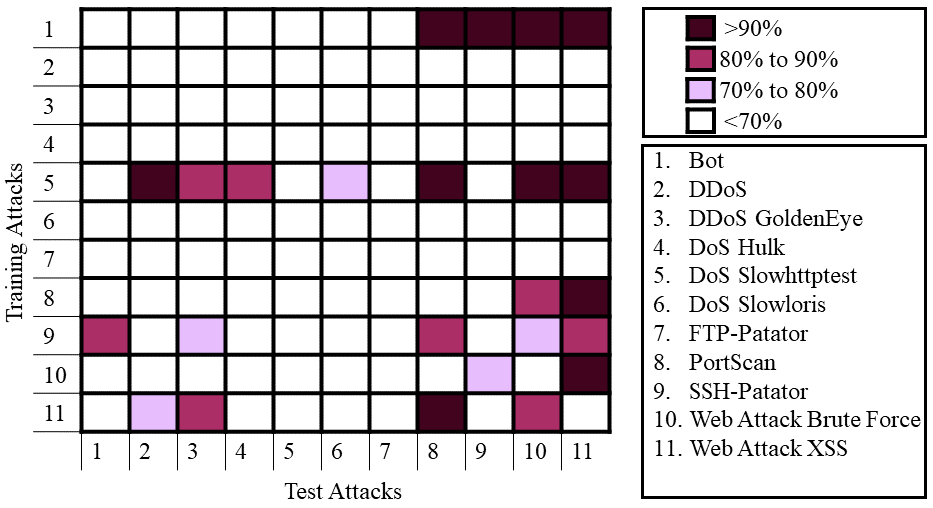}
    \caption{Transferability Pairs Detected by TabFIDSv2 with a Autoencoder Backend (25-10-10 Train-Validation-Test Split).}
    \label{fig:autoencoder_tabfids_v2_25}
    \vspace{-8pt}
\end{figure}

The results from this approach are presented in Fig. \ref{fig:autoencoder_tabfids_v2_25}, with a 25-10-10 split. We get 24 transferable pairs. One interesting point to note here is that 5 training attacks and 2 test attacks are absent from any transferable pairs. In the original TabFIDSv2 or the Resnet backend, every train or test attack was represented in multiple (train,test) pairs. This indicates the importance of choosing a backend carefully.

Finally, in Table \ref{table:train-test-stats}, we summarize the notable experimental results. Here, we note the total number of transferable pairs achieved by each approach, as well as a breakdown of very high ($\geq$90\%), high (80-90\%), and moderate (70-80\%) transferability. We also include the localized intrusion detection rates for each of these approaches. We note that both TabFIDS and TabFIDSv2 achieve excellent transferability while maintaining very high localized intrusion detection rates. The combination of TabFIDSv2, with a 25-10-10 train-validation-test split, and a temporal windows of size 5 achieves the best transferability. While the Resnet backend performs pretty well, the autoencoder backend underperforms in both transferability and localized intrusion detection.

\subsection{TabFIDSv2 with a different dataset}

\begin{figure}[t]
    \centering
    \includegraphics[width=0.45\textwidth]{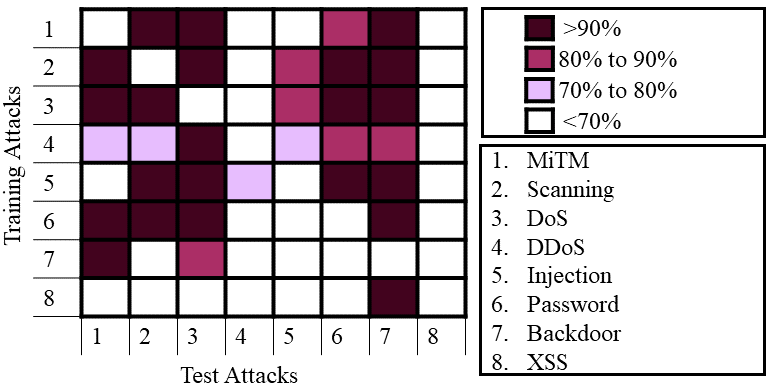}
    \caption{Transferability Relationships with Tabfidsv2 on the Ton-IoT Dataset.}
    \label{fig:tabfids_v2_ton_iot}
\end{figure}

To further investigate the generalizability of TabFIDSv2, we replace the  CIC-IDS 2017 dataset with the CIC-ToN\_IoT dataset\cite{toniot}. This analysis aims to show that the versatility of the proposed approach is not dependent on a specific dataset. The CIC-ToN\_IoT dataset consists of benign data along with 8 different types of attacks-XSS, Injection, Password, Scanning, MITM, DDoS, DoS, and Backdoor. 

For this experiment, the non-numeric features are removed from the dataset. Similar to the approach in \cite{ruzafa2021intrusion}, we restrict our analysis to the data samples corresponding to the top 10 destination IP addresses. This data is then distributed among 8 nodes. The benign data packets are divided equally to each of the 8 nodes. Each node gets only one class of attack data. This distribution is similar to the the one followed earlier for CIC-IDS 2017. The results for TabFIDSv2 on this dataset are presented in Fig. \ref{fig:tabfids_v2_ton_iot}. The train, validation and test split for this experiment is 80-10-10. Here, we notice transferability in 32 pairs, out of a possible 56 pairs. Out of these 32 pairs, 22 pairs show transferability with >90\%. This performance validates the generalizability of our approach in uncovering transferability across a wide variety of attacks. 

\section{Conclusion}

In this paper, we develop TabFIDSv2, a novel intrusion detection system designed to optimize transferability in federated learning environments. By integrating two pre-processing techniques\textemdash bootstrapping and temporal averaging\textemdash with our empirically developed deep neural network, our proposed TabFIDSv1 algorithm achieves exceptional localized detection performance and high transferability across significantly imbalanced datasets. We also evaluate a Data-Driven Feature Elimination algorithm that presents a reasonable trade-off between transferability and processing cost. The integration of a Block-Based Smart Aggregation algorithm, resulting in TabFIDSv2, further enhances transferability performance and localized intrusion detection accuracy. Comprehensive evaluations demonstrate the system's generalizability across diverse neural network backbones, and different datasets. Although TabFIDSv2 is developed with a focus on intrusion detection systems, this development can be of value in other deep learning domains where transferability is a desired attribute.

\ifCLASSOPTIONcaptionsoff
  \newpage
\fi
\bibliographystyle{IEEEtran} 

\bibliography{nattack2021}

\end{document}